\documentstyle[11pt,iau185,twoside,epsf]{article}

\begin{document}
\title{Nonradial Pulsations in Classical Pulsators}
\author{G. Kov\'acs}
\affil{Konkoly Observatory, P.O. Box 67, H-1525 Budapest, Hungary}
%%%%%\altaffiltext{1}{e-mail: kovacs@konkoly.hu}

%%%%%%%%%%%%%
%           %
% ABSTRACT  %
%           %
%%%%%%%%%%%%%

\begin{abstract}
Recent analyses of photometric data on globular clusters and galaxies 
enabled us to study more closely the long-periodic amplitude/phase 
modulation (Blazhko effect) in classical variables. In the frequency 
spectra of these stars we see either a doublet or an equally-spaced 
triplet with a very small frequency separation close to the main 
component. None of the available theoretical models are able to 
explain this behavior without invoking some form of nonradial 
pulsation. In this review we describe the observational status of the 
Blazhko variables, and discuss the limits of the applicability of the 
current models to these stars.     
\end{abstract}


%%%%%%%%%%%%%
%           %
%  SECT. 1  %
%           %
%%%%%%%%%%%%%
 
\section{Introduction}
The long-term periodic phase modulation of the \object{RR\,Lyrae} 
star \object{RW\,Dra} was discovered almost one hundred years ago 
by Blazhko (1907). Therefore, this phenomenon is called 
{\it Blazhko effect} (for the nomenclature see also the comment of 
Szeidl \& Koll\'ath 2000). 

Long-term amplitude/phase modulation is quite common in the case of 
multimode pulsators (e.g., $\delta$~Scuti stars, white dwarfs), where 
the effect of close excited normal modes can also be viewed as an 
amplitude modulation. What makes Blazhko-type stars particularly 
interesting and important is that they have high pulsation amplitudes, 
and therefore there is no doubt that their main pulsation 
component is basically a radial mode. Since the long-term 
amplitude/phase modulations are exhibited in the frequency spectra as 
closely spaced components, and the eigenspectrum of the radial modes 
is too sparse, we are left with the following two possibilities to 
explain the Blazhko effect:
\begin{itemize}
\item
amplitude/phase-modulated pulsation of a purely radial mode,
\item
basically radial pulsation, contaminated by nonradial component(s) 
either through mode coupling, or through the distortion of the 
radial eigenfunction by a large-scale magnetic field.  
\end{itemize}   
Convection can be considered also in the above scheme, because (at least 
in the first approximation) it can be treated as a large set of nonradial 
modes of high spherical harmonic order. Of course, if any form of 
convection is responsible for the amplitude modulation, the interaction 
with the radial normal mode should lead to a genuine amplitude modulation 
which is independent of the aspect angle, because the averaging effect 
is certainly overwhelming in this case. 

In the following we will give an account of the observed properties of 
the Blazhko variables and describe the ability of the current models to 
explain the observations. We will see that, according to our current 
understanding, it is very difficult to escape the possibility that 
nonradial pulsations play a significant role in these variables.

%%%%%%%%%%%%%
%           %
%  SECT. 2  %
%           %
%%%%%%%%%%%%%
 
\section{Data analysis}
Because of the long modulation periods, identification of Blazhko 
variables requires sufficiently long data coverage and proper 
analysis. While in the case of old photographic data the high noise 
level is the most serious problem, the photoelectric observations,  
made on individual variables during the past decades, suffer from 
another problem. Besides the daily alias in the sampling of these 
data there is another artificial periodicity due to the high 
concentration of data points on the rising branches. This observation 
technique was justified because of the input data required by the 
$O-C$ analyses, most frequently used in the past to study these 
stars. The sampling periodicity with the pulsation period makes it 
very difficult/impossible to perform a Fourier analysis and correctly 
identify the modulation components around the harmonics. 

Data supplied by the microlensing surveys and observations made on 
individual globular clusters suffer from problems, too. 
Microlensing data have rather low sampling rate and usually high 
noise (at least at the brightness level of the RR~Lyrae stars). 
Although individual cluster data have much higher sampling rate, 
and, in general, are more accurate, they are far less extended. 
Nevertheless, analysis of the above types of data during the last 
few years has led to significant progress in the study of these 
stars.     

As an example of the data quality and types of frequency patterns 
obtained in the recent analysis of the {\sc macho} database on the 
Large Magellanic Cloud (LMC) fundamental mode RR~Lyrae (RRab) stars 
(Welch et al. 2002), we show the result for five representative 
variables in Fig.~1. 
%
%
%>>>>>>>>>>>>> FIG. 1
%
\begin{figure}
\begin{center}
\mbox{
\epsfxsize=0.8\textwidth\epsfysize=0.8\textwidth\epsfbox{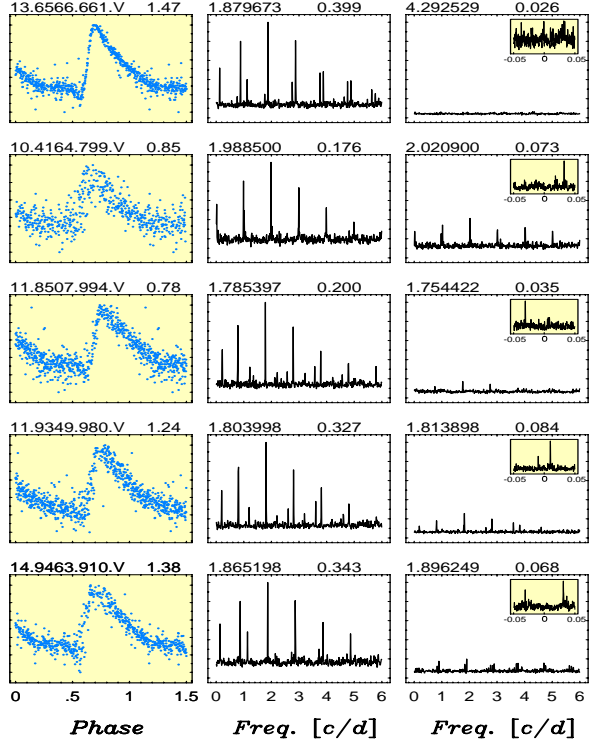}}
\caption{Selected samples of light curves and frequency spectra of 
the {\sc macho} fundamental mode RR~Lyrae star inventory in the LMC. 
Each row represents different objects as given by the first item on 
the left in the corresponding header. From left to right we have: 
\underline {first column:} folded light curves, 
\underline {second column:} amplitude spectra of the original data, 
\underline {third column:} amplitude spectra of the prewhitened data, and  
\underline {insets in the upper right corners:} blow-ups of the 
neighborhood of the fundamental frequency in the prewhitened spectra. 
Light curves, frequency spectra of the original data and those in the 
insets are normalized independently. Prewhitened spectra are normalized 
by the highest amplitude in the corresponding spectra of the original 
data. In the headers from left to right are shown: {\sc macho} identifier, 
total $V$ amplitude, fundamental frequency [d$^{-1}$] and its amplitude, 
peak frequency and its amplitude in the prewhitened spectra.}   
\end{center}
\end{figure}
The top panel is only to demonstrate that for single mode variables, 
the prewhitened spectrum is completely featureless, presuming that 
sufficient number of harmonics (in this case three) are subtracted 
from the signal. In the next two panels from the top we show cases 
when only one side component is seen after prewhitening, whereas the 
lower two panels are examples when both modulation components are 
clearly visible. These types of frequency spectra have also been 
observed in the first overtone RR~Lyrae (RRc) stars (see Alcock et al. 
2000; Cseresnjes 2001; Olech et al. 2001, and references therein). 
In this paper variables displaying {\it any} of the above types of 
frequency patterns are called Blazhko (BL) variables, because at 
the time of this writing it is not entirely clear if there is an 
abrupt or continuous transition between the cases of strongly 
asymmetric and single modulation components. 

Although additional analyses of new and already existing data will 
continuously increase the number of Blazhko stars with reliably 
identified frequency patterns, we think it is useful to present the 
number statistics available for us at this moment. Several comments 
should be added to the numbers displayed in Table~1. 
%
%
%>>>>>>>>>>>>> TABLE 1
%
\begin{table}
\caption{Number of known Blazhko variables in various stellar systems} 
\tabcolsep=4pt
\begin{center}
\begin{tabular}{lrrl}
\hline\\[-10pt] Cluster/Galaxy & $N_{RRab-BL}$ & $N_{RRc-BL}$ & Source  \\
\hline\hline\\[-10pt]
Galactic field     &  37\phantom{-----} &  8\phantom{-----} &  1, 5     \\
Galactic Bulge     &  35\phantom{-----} &  2\phantom{-----} &  6        \\
Globular clusters  &   ?\phantom{-----} &  7\phantom{-----} &  2, 3, 4  \\
LMC                & 149\phantom{-----} & 52\phantom{-----} &  7, 8     \\
Sgr dwarf gal.     &   ?\phantom{-----} &  5\phantom{-----} &  5        \\
\hline\\[-5pt]
\end{tabular}
\parbox{82mm}{
{\footnotesize
\underline {References:} (1) Szeidl (1988); (2) Olech et al. (1999a);   
(3) Olech et al. (1999b); (4) Olech et al. (2001); (5) Cseresnjes, P. 
(2001); (6) Moskalik \& Poretti (2002); (7) Alcock et al. (2000); 
(8) Welch et al. (2002)} 
\\
{\footnotesize
\underline {Note:} See text for the discussion of the incompleteness 
of the above statistics.}
}
\end{center}
\end{table}
For the RRab-BL stars in the Galactic field, we rely on the review of 
Szeidl (1988), who listed `RRab stars with known or presumed Blazhko 
period' in his Table~3. We note that there are stars (mentioned 
also by Szeidl) which might not be Blazhko variables, because they do 
not show definitive change in the shape of their light curves in the 
contemporary observations, whereas the old photographic data might 
indicate such changes. Furthermore, partially because of the problems 
due to the biased sampling as mentioned at the beginning of this section, 
only a few of these stars have been frequency analyzed (Borkowski 1980; 
Smith et al. 1994; Kov\'acs 1995; Nagy 1998; Smith et al. 1999; 
Szeidl \& Koll\'ath 2000; Lee \& Schmidt 2001, and references therein). 

In the case of globular clusters, we included only those variables 
which have been discovered from the frequency analysis of current 
{\sc ccd} observations. Unfortunately, none of these analyses were 
extended to RRab stars. Therefore, we did not include any data for 
the RRab-BL stars. For these variables we refer the interested reader 
to the statistics based mostly on the visual inspection of photographic 
data as presented by Szeidl (1988). 

We also do not have complete information on the statistics of the 
Blazhko phenomenon for the Sagittarius dwarf galaxy. This is because 
the analysis of the RRab stars was dropped due to the short time span 
of the observations. 

Periodic amplitude/phase variation among Cepheids must be very rare. 
The only known Blazhko variable in the Galactic field classified as 
a Cepheid, is \object{HR\,7308} (Burki et al. 1986; see also Van 
Hoolst \& Waelkens 1995). A preliminary analysis of the {\sc macho} 
database of the Magellanic Clouds suggests an incidence rate of 
$0.5-1.0$\% (Welch, 2001, private communication). Our own analysis 
of some 1400 Cepheids in the same system from the {\sc ogle} database 
(Udalski et al. 1999), has led basically to a null result. Considering 
that the Blazhko Cepheids identified so far have short periods and our 
analysis covered a wide range of periods, these results seem to 
suggest an incidence rate lower than 1\% for these variables. This 
is an order of magnitude lower than the recent estimate of the 
incidence rate of the Blazhko RR~Lyrae stars in the LMC (Welch et al. 
2002).

%%%%%%%%%%%%%
%           %
%  SECT. 3  %
%           %
%%%%%%%%%%%%%
 
\section{Observed properties}
As we mentioned in the previous section, the statistics of the Blazhko 
stars is incomplete in many systems. The only exception is the LMC, 
where, due to the analysis of several thousands stars, we have a sample 
of $\approx 200$ Blazhko stars (RRab \& RRc). Based mainly on these stars,
we summarize the properties obtained from frequency analyses in Fig.~2. 
%
%
%>>>>>>>>>>>>> FIG. 2
%
\begin{figure}[t]
\begin{center}
\vskip 0mm
\mbox{\epsfxsize=0.4\textwidth\epsfysize=0.4\textwidth\epsfbox{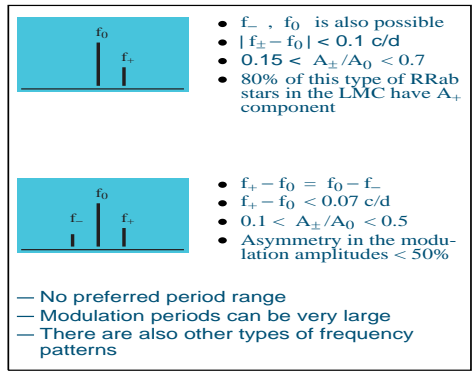}}
\vskip -5mm  
\caption{Main properties of the Blazhko variables}
\vskip -5mm   
\end{center}
\end{figure}

First of all, we have to point out that, although the data strongly 
suggest the existence of two main classes, the noise level is 
still too high to separate clearly the class of variables with two 
modulation components (BL2-type) from the ones which apparently have 
only one such a component (BL1-type). Significant doublets rarely 
appear with an asymmetry larger than 50\% in the prewhitened spectra 
of the BL2 variables. However, less significant peaks are sometimes 
observable at the equally spaced frequency position in the case of 
variables classified as BL1-type. Even though we cannot exclude the 
existence of BL2 variables with strongly asymmetric modulation 
components, we think that it is safe to state that in the large 
majority of BL2 stars this asymmetry cannot be larger than 70\%. 

Modulation periods are larger than $\approx 20$~days for the RRab 
stars whereas for the RRc stars they can be even shorter than 10~days. 
For the upper limit of the modulation periods we may not have such a  
relatively firm statement. The longest modulation periods observed in 
the {\sc macho} database are comparable with the length of the total 
time span of the observations, which is 7.5~years. The modulation 
periods for the RRc-BL stars are distributed roughly uniformly with 
some preference toward longer periods. For the RRab-BL stars, modulation   
periods shorter than $\approx 40$~days are not frequent, but above this 
value the distribution is reasonably uniform. 

The pulsation periods cover almost the total range of that of the 
monoperiodic variables. For the RRab- and RRc-BL stars we have 
$0\fd35<P_0<0\fd7$ and $0\fd23<P_0<0\fd46$, respectively.

There are a few RRc-BL stars with very strong modulation 
($A_{\pm}/A_0>0.7$). In about two third of the BL2 stars (both among the 
RRab and RRc stars) the modulation component with the larger frequency  
has also larger amplitude. In 80\% of the RRab-BL1 variables the 
modulation component has larger frequency than the pulsation component. 
There is no such trend among the RRc-BL1 stars. 

To estimate the incidence rate of the Blazhko phenomenon, we turn to 
the {\sc macho} data set which contains enough variables to yield 
reliable figures. According to Alcock et al. (2000) and Welch et al. 
(2002), the incidence rates of the BL1 \& BL2 phenomena in the LMC are  
about 10\% and 4\% among the RRab and RRc variables, respectively. 
Based on the analysis of 214 stars from the {\sc ogle} data set, 
Moskalik \& Poretti (2002) get rates of 23\% and 3\% for the Galactic 
Bulge. We think that these results clearly show that, in general, the 
Blazhko phenomenon has at least three times higher incidence rate 
among RRab stars than among RRc stars. However, the actual value of 
this ratio depends on the system studied.

%%%%%%%%%%%%%
%           %
%  SECT. 4  %
%           %
%%%%%%%%%%%%%
 
\section{Physical modelling}
The reasons why all current models include some form of nonradial 
pulsations in explaining Blazhko effect are the following:
\begin{itemize}
\item
Previous attempts in using only radial mode interactions have failed 
(nonresonant modes -- Buchler \& Kov\'acs 1986;
2:1 resonance -- Moskalik 1986;
noise induced transitions -- Kov\'acs 1994;
no reports from past and current hydrodynamical simulations -- 
Koll\'ath 2001, private communication).
\item
Linear models suggest excitation of nonradial modes 
(Cepheids -- Osaki 1977; 
RR~Lyrae stars -- Dziembowski 1977; Cox 1993; Dziembowski \& Cassisi 1999;  
beyond the blue edge -- Shibahashi \& Osaki 1981).
\item
Observed frequency patterns are compatible with the assumption of 
nonradial pulsation (equidistant triplet, closely spaced frequencies).
\end{itemize} 
%
%
%>>>>>>>>>>>>> FIG. 3
%
\begin{figure}[t]
\begin{center}
\vskip 0mm
\mbox{\epsfxsize=0.4\textwidth\epsfysize=0.4\textwidth\epsfbox{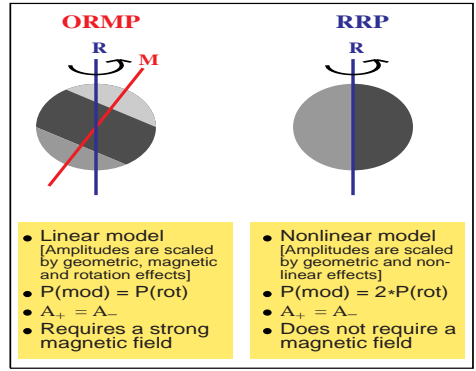}}
\vskip -5mm  
\caption{Comparison of two models of the Blazhko effect.}
\vskip -5mm   
\end{center}
\end{figure}
We show the schematic representation of the two currently competing 
models in Fig.~3. Both in the Oblique Rotating Magnetic Pulsator (ORMP, 
see Shibahashi 2000) and in the Rotating Resonant Pulsator (RRP, see 
Nowakowski \& Dziembowski 2001) models the pulsation amplitudes are 
{\it constant}. The observed modulation of the light curve is a 
consequence of {\it rotation}, and the modulation period is directly 
related to it. The degree of modulation is strongly 
{\it aspect-dependent} in both cases. Another peculiarity of these 
models is that they predict modulation components of {\it equal 
amplitudes}, which is in sharp contrast to most of the frequency 
patterns observed, especially with those which are of BL1-type. 
A further question to deal with is the reason why RRc-BL variables 
show significantly lower incidence rates than their fundamental 
mode counterparts. This question has not yet been dealt with in the 
ORMP model, but the RRP model does show such an effect, although 
still not in the degree observed. Finally, it is important to mention 
that deviations from strict amplitude/phase modulations might occur 
in Blazhko stars (e.g., Szeidl 1988; Smith et al. 2002) which should 
be explained by future modelling.

%%%%%%%%%%%%%
%           %
%  SECT. 5  %
%           %
%%%%%%%%%%%%%
 
\section{Problems and prospects}
In spite of the progress made during the last few years both in the 
observations and modelling, the basic physical understanding of the 
Blazhko phenomenon is still missing. Posed by the observations, here 
is a (probably incomplete) list of problems to be dealt with: 
-- uneven modulation components; -- incidence rates (high for RRab, 
much lower for RRc and almost zero for Cepheids); -- deviations from 
strictly periodic modulations; -- magnetic field (spectroscopic 
detection and its relation to the Blazhko effect); -- surface velocity 
field (spectroscopic detection of the nonradial component). Present 
models are still not the results of direct hydrodynamical simulations, 
but rely either on linear pulsation (OMRP) or on the amplitude equation 
formalism, using some numerical estimates of the coupling coefficients. 
Further progress in this field requires substantial efforts. From 
the observational side, disentangling the radial and nonradial components 
of the surface velocity field and detecting/studying magnetic field 
would be of prime importance. From the theoretical side, a full scale 
hydrodynamical simulation is still a formidable task. Therefore, 
additional work within the framework of amplitude equations seems 
to be the most promising. Considering the fundamental importance 
of RR~Lyrae stars in various fields of astronomy, we think that it is 
very much worthwhile to put substantial efforts in understanding this 
long-standing problem. 

\pagebreak

\acknowledgments
Useful discussions with Conny Aerts, Wojtek Dziembowski, Doug Welch, 
Pawel Moskalik, Rafal Nowakowski and Arkadiusz Olech are very much 
appreciated. We thank the organizers of this meeting for the financial 
help and hospitality. This work was completed during the 
author's visit to the Copernicus Astronomical Center, Warsaw. The 
following grants are acknowledged: {\sc otka} {\sc t--026031},  
{\sc t--030954} and {\sc t--038437}.

\end{document}